\documentclass[12pt,journal,final,onecolumn]{IEEEtran} 

\usepackage[pdftex]{graphicx}
\usepackage[cmex10]{amsmath}
\usepackage{amssymb}
\usepackage{amsthm}
\usepackage{amsfonts}
\usepackage{mathtools} 
\usepackage{thmtools} 
\usepackage{bm}
\usepackage{cite}
\usepackage[svgnames]{xcolor} 
\usepackage{pstricks,pst-node,pst-plot,pstricks-add}
\usepackage[tight,footnotesize]{subfigure}
\usepackage[binary-units=true]{siunitx}
\usepackage{hyperref} 

\graphicspath{{figs/}}

\mathtoolsset{showonlyrefs} 

\newcommand{\mc}[1]{\mathcal{#1}}

\newcommand{\msf}[1]{\mathsf{#1}}

\newcommand{\defeq}{\mathrel{\triangleq}}
\newcommand{\Pp}{\mathbb{P}}
\newcommand{\E}{\mathbb{E}}
\newcommand{\N}{\mathbb{N}}
\newcommand{\Z}{\mathbb{Z}}

\newcommand{\R}{\mathbb{R}}

\DeclarePairedDelimiter\abs{\lvert}{\rvert}

\DeclarePairedDelimiter\norm{\lVert}{\rVert}
\DeclarePairedDelimiter\bracket{\langle}{\rangle}

\newcommand{\iid}{i.\@i.\@d.\ }
\newcommand{\T}{\msf{T}}

\newcommand{\semcol}{;}

\DeclareMathOperator{\tr}{tr}

\newtheorem{lemma}{Lemma}

\newtheorem{theorem}[lemma]{Theorem}

\newtheorem{innercustomtheorem}{Theorem}
{\renewcommand\theinnercustomtheorem{#1}\innercustomtheorem}%
{\endinnercustomtheorem}

\theoremstyle{definition}

\renewcommand\thmcontinues[1]{Continued}

\newtheoremstyle{myremark}%
{\topsep}{\topsep}{\normalfont}{\parindent}{\itshape}{:}{ }{}

\theoremstyle{myremark}

\newcommand{\dop}{\msf{DOP}}

\begin{document}

\title{Satellite Positioning with Large Constellations} 

\author{Urs Niesen and Olivier L{\'e}v{\^e}que%
    \thanks{U. Niesen is with the Qualcomm New Jersey Research Center, Bridgewater, NJ 08807, USA.  O. L{\'e}v{\^e}que is with EPFL, 1015 Lausanne, Switzerland.  Emails: urs.niesen@ieee.org, olivier.leveque@epfl.ch}%
}

\maketitle

\begin{abstract}
    Modern global navigation satellite system receivers can access signals from
    several satellite constellations (including GPS, GLONASS, Galileo, BeiDou).
    Once these constellations are all fully operational, a typical receiver can
    expect to have on the order of $40$--$50$ satellites in view. Motivated by
    that observation, this paper presents an asymptotic analysis of positioning
    algorithms in the large-constellation regime. We determine the exact
    asymptotic behavior for both pseudo-range and carrier-phase positioning. One
    interesting insight from our analysis is that the standard carrier-phase
    positioning approach based on resolving the carrier-phase integer
    ambiguities fails for large satellite constellations. Instead, we adopt a
    Bayesian approach, in which the ambiguities are treated as noise terms and
    not explicitly estimated.
\end{abstract}

\section{Introduction} 
\label{sec:intro}

\subsection{Motivation and Summary of Results}
\label{sec:intro_motivation_summary}

In order to determine its position and clock bias, a global navigation satellite
system (GNSS) user needs to have at least four satellites in view. A fully
operational GNSS constellation guarantees that this condition is always
satisfied (in open sky), and traditional positioning algorithms were designed
and analyzed with a number of visible satellites on that order in mind. 

However, modern GNSS receivers can access signals from several different GNSS
constellations including GPS (the US system), GLONASS (Russian), Galileo
(European), and BeiDou (Chinese). Once these are fully deployed (which is
already the case for GPS and GLONASS, and is expected by the end of this decade
for Galileo and BeiDou), each of these constellations will have around $30$
operational satellites. Thus, combined, there will be around $120$ operational
GNSS satellites~\cite{li15}. A typical receiver in open-sky condition will then
have access to on the order of $40$--$50$ visible satellites.

This large number of visible satellites motivates an asymptotic analysis of the
performance of GNSS positioning algorithms. We consider two different types of
positioning approaches. The first approach uses only pseudo-range measurements.
The second approach uses in addition carrier-phase measurements. We provide an
asymptotic analysis of both these positioning approaches in the
large-constellation regime.

When only pseudo-range measurements are available, the maximum likelihood (ML)
estimate of the position is equal to the least-squares (LS) estimate. Its
positioning performance depends on the satellite geometry and is summarized by
the so-called dilution of precision (DOP). In order to analyze the performance
behavior asymptotically, we introduce a simple stochastic model for the
distribution of satellites across the sky. Using this model,
Theorem~\ref{thm:pr} below shows that the DOP decreases as the inverse of the
square root of the number of visible satellites and provides the exact scaling
constant in front of the square-root term.

When carrier-phase measurements are also available, a more accurate positioning
is possible. Unfortunately, the carrier-phase measurements are corrupted by an
unknown integer ambiguity. The standard approach is to explicitly estimate these
ambiguities as nuisance parameters. State-of-the-art estimation methods for
resolving the ambiguities efficiently are LAMBDA~\cite{teunissen95} and
modified LAMBDA~\cite{chang05}.  For moderate number of visible
satellites, these methods allow to resolve all ambiguities, leading to
significantly improved performance compared to pseudo-range only positioning.
However, for large number of visible satellites, insisting on resolving the
integer ambiguities of all carrier-phase measurements leads to resolution errors
for at least some of them, which deteriorates the positioning performance. A
different treatment of these integer ambiguities is therefore required.

In the present paper, we instead adopt a Bayesian approach, treating the integer
ambiguities as noise. This leads to an interesting expression for the
maximum-likelihood estimate of the position (Equation~\eqref{eq:cpml} below),
involving the minimum mean-squared error (MMSE) estimates of the integer
ambiguities.  We characterize the asymptotic behaviour of this ML estimate in
Theorem~\ref{thm:cp} below. The standard deviation of the ML estimate is shown
to decrease as the inverse of the square root of the number of satellites.
Furthermore, the usefulness of the carrier-phase measurements is characterized
by the ratio between the carrier wavelength and the carrier-phase noise standard
deviation. More precisely, it is shown that the (rescaled) variance of the ML
estimate can be expressed as a function that solely depends on this ratio.

\subsection{Related Work}
\label{sec:intro_related}

There has been significant recent interest in multi-constellation positioning,
with performance evaluations both through simulations~\cite{wang12, li15} and
experiments~\cite{cai13, montenbruck14, uhlemann15}. These results indicate that
the many satellites available in these multi-constellation systems can lead to
lower and less variable DOP~\cite{li15}, shorter convergence time of
positioning algorithms~\cite{cai13}, and better service availability in urban
scenarios~\cite{wang12}.

The traditional method used for carrier-phase positioning is to resolve integer
ambiguities prior to estimating the baseline coordinates. Multiple methods have
been developed along these lines, including the works of
Teunissen~\cite{teunissen95} and Chang et al.~\cite{chang05} (already mentioned
above), Hatch~\cite{hatch90}, Remondi~\cite{remondi91}, Al-Haifi et
al.~\cite{alhaifi98}, and Hassibi and Boyd~\cite{hassibi98}. Recent interest in
positioning with multiple satellite constellations has spurred the development
of integer ambiguity resolution algorithms that scale more favorably with the
number of satellites~\cite{cellmer18}.

Parallel to this line of works, various authors have proposed following a
Bayesian approach in order to estimate the baseline coordinates, treating
integer ambiguities as noise and not imposing their resolution prior to the
coordinates' estimation~\cite{blewitt89, betti93, gundlich02}. Building on these
approaches, de Lacy et al.~have proposed in~\cite{delacy02} to use Monte-Carlo
simulations in order to refine the search space for the integer ambiguities.
Also using a Bayesian framework, Garcia et al.~have proposed in~\cite{garcia16}
an adaptive method to refine the estimation of the baseline coordinates when new
measurements become available.

\subsection{Organization}
\label{sec:intro_organization}

The remainder of this paper is organized as follows. Section~\ref{sec:setting}
formally introduces the problem setting. Section~\ref{sec:results} presents the
main results. Section~\ref{sec:conclusion} contains concluding remarks. All
proofs are deferred to the appendix.

\section{Problem Setting}
\label{sec:setting}

We consider the GNSS positioning problem with $S$ satellites, where $S$ is
assumed large. For each satellite $s\in\{1, 2, \dots, S\}$, we obtain two
measurements, a pseudo-range measurement and a carrier-phase measurement. 

After appropriate linearization around an approximate position solution and
subtraction of known terms, the pseudo-range $\msf{y}_s$ satisfies the (approximately)
linear measurement equation
\begin{equation}
    \label{eq:pr_meas}
    \msf{y}_s \defeq -\bm{u}_s^\T\bm{x}+b+\sigma \msf{z}_s.
\end{equation}
See, e.g., \cite[Chapter~6.1.1]{misra06} for a detailed derivation.
Here, $\bm{x}\in\R^3$ is the receiver position (technically the position
approximation error), $b\in\R$ is the receiver clock bias (again, technically
the clock bias approximation error), $\bm{u}_s$ is the unit vector from the
receiver to satellite $s$ (computed from the approximate position solution), and
$\sigma \msf{z}_s$ is receiver noise. This receiver noise is assumed here to be a
Gaussian random variable with mean zero and variance $\sigma^2$, and to be
independent and identically distributed (i.i.d.) across satellites. Observe that here and
in the following we use sans-serif font (i.e., $\msf{y}_s$, $\msf{z}_s$) to
indicate random quantities.

It will be convenient to define the $S$-dimensional vector of pseudo-range
measurements
\begin{equation*}
    \bm{\msf{y}} \defeq \bigl( \msf{y}_s \bigr)_{s=1}^S
\end{equation*}
and similar for $\bm{\msf{z}}$. Further, define the $S\times 4$ design matrix
\begin{equation}
    \label{eq:Gdef}
    \bm{G} \defeq 
    \begin{pmatrix}
        -\bm{u}_1^\T & 1 \\
        -\bm{u}_2^\T & 1 \\
        \vdots & \vdots \\
        -\bm{u}_S^\T & 1
    \end{pmatrix}.
\end{equation}
and the $4 \times 1$ vector of unknown parameters
\begin{equation} 
    \label{eq:wdef}
    \bm{w} \defeq \begin{pmatrix} \bm{x} \\ b \end{pmatrix}
\end{equation}
With these definitions, the pseudo-range measurement equation~\eqref{eq:pr_meas}
can be rewritten in vector form as
\begin{equation}
    \label{eq:pr_measv}
    \bm{\msf{y}} = \bm{G} \bm{w} + \sigma \bm{\msf{z}}.
\end{equation}

Similarly, the carrier phase $\tilde{\msf{y}}_s$ can be linearized to satisfy
the approximate measurement equation
\begin{equation}
    \label{eq:cp_meas}
    \tilde{\msf{y}}_s \defeq -\bm{u}_s^\T\bm{x}+b+\lambda m_s+\tilde{\sigma} \tilde{\msf{z}}_s;
\end{equation}
see, e.g., \cite[Chapter~7]{misra06}.  Here $\lambda$ is the carrier wavelength
(around $\SI{0.19}{\meter}$ for the GPS L1 signal), $m_s\in\Z$ is the
unknown integer ambiguity, and $\tilde{\sigma} \tilde{\msf{z}}_s$ is receiver noise. This
receiver noise is assumed to be Gaussian with mean zero and variance
$\tilde{\sigma}^2$, \iid across satellites and independent of $\bm{\msf{z}}$. 

We can again define the $S$-dimensional vector of carrier-phase measurements
\begin{equation*}
    \tilde{\bm{\msf{y}}} \defeq \bigl( \tilde{\msf{y}}_s \bigr)_{s=1}^S,
\end{equation*}
and similar for $\bm{m}$ and $\tilde{\bm{\msf{z}}}$.
The carrier-phase measurement equation~\eqref{eq:cp_meas} then becomes
\begin{equation}
    \label{eq:cp_measv}
    \tilde{\bm{\msf{y}}} 
    = \bm{G} \bm{w} + \lambda\bm{m}+\tilde{\sigma} \tilde{\bm{\msf{z}}}
\end{equation}
with $\bm{G}$ as defined in~\eqref{eq:Gdef}.

The carrier-phase measurements are much more precise than the pseudo-ranges.
Typically, $\tilde{\sigma}$ is around a factor $100$ smaller than $\sigma$ (see,
e.g., \cite[Chapter~5.5]{misra06}).  However, the carrier phases have the
disadvantage that they contain an unknown integer ambiguity. Dealing with these
integer ambiguities is one of the key challenges in carrier-phase positioning.

The above measurement model captures only first-order effects. In particular,
atmospheric and ephemeris errors are neglected. Thus, this model is appropriate
assuming that those errors have been corrected, for example using differential
corrections.

The measurement equations \eqref{eq:pr_meas} and \eqref{eq:cp_meas} are stated
for known, deterministic satellite positions (captured by the unit vectors
$\bm{u}_s$).  To enable analytical evaluations of the positioning performance,
we require a model for these unit vectors. We assume in the following that each
$\bm{\msf{u}}_s$ is independently and uniformly distributed over the (say
northern) hemisphere. This somewhat stylized model allows for analytical
tractability and does again capture the first-order behavior. 

With this assumption, the unit vector $\bm{\msf{u}}_s$ is now a random variable.
As a consequence, the design matrix $\bm{\msf{G}}$ defined in~\eqref{eq:Gdef} is
now also a random matrix.  The problem considered throughout the remainder of
this paper is thus to estimate the receiver position $\bm{x}$ and clock bias $b$
from the pseudo-ranges $\bm{\msf{y}}$, the carrier phases
$\tilde{\bm{\msf{y}}}$, and the satellite unit vectors $\bm{\msf{u}}_1, \dots,
\bm{\msf{u}}_S$ (or, equivalently, the design matrix $\bm{\msf{G}}$). In
particular, we will be interested in the estimation performance as the number of
satellites $S$ increases.

\section{Main Results}
\label{sec:results}

We start with an analysis of pseudo-range only positioning in
Section~\ref{sec:results_pr}. This will lay the foundation for our discussion of
carrier-phase positioning in Section~\ref{sec:results_cp}.

\subsection{Pseudo-Range Positioning}
\label{sec:results_pr}

The ML estimator of the parameter vector $\bm{w}$ (consisting of the receiver
position $\bm{x}$ and clock bias $b$) given the pseudo-range measurement vector
$\bm{\msf{y}}$ and the design matrix $\bm{\msf{G}}$ is
\begin{equation}
    \label{eq:prml}
    (\bm{\msf{G}}^\T\bm{\msf{G}})^{-1}\bm{\msf{G}}^\T\bm{\msf{y}}
\end{equation}
This estimator is easily seen to be Gaussian with mean $\bm{w}$ and covariance
matrix
\begin{equation}
    \label{eq:prcov}
    \sigma^2(\bm{\msf{G}}^\T\bm{\msf{G}})^{-1}
\end{equation}
(see, e.g., \cite[Chapter~6.1]{misra06}).

The quality of the estimate~\eqref{eq:prml} depends on the satellite geometry
through the design matrix $\bm{\msf{G}}$. This dependence is often summarized
into a scalar quantity called the (geometric) dilution of precision, defined as
\begin{equation*}
    \dop(\bm{\msf{G}}) 
    \defeq \sqrt{\tr\bigl((\bm{\msf{G}}^\T\bm{\msf{G}})^{-1}\bigr)}
\end{equation*}
(see again \cite[Chapter~6.1]{misra06}). Here $\tr(\cdot)$ denotes the trace.
Observe that $\dop(\bm{\msf{G}})$ is a random variable due to the random nature
of $\bm{\msf{G}}$.

The $\dop$ is lowest (and hence estimation performance best) if the satellites
are well distributed across the hemisphere. For small number of visible
satellites, the $\dop$ can vary quite significantly. However, as our first
theorem shows, this variability reduces as the number of satellites increases.

\begin{theorem}
    \label{thm:pr}
    The scaled pseudo-range positioning covariance matrix
    $S\cdot\sigma^2(\bm{\msf{G}}^\T\bm{\msf{G}})^{-1}$ converges in probability
    to $\sigma^2\bm{Q}$ with
    \begin{equation*}
        \bm{Q} \defeq
        \begin{pmatrix}
            3 & 0 & 0 & 0 \\
            0 & 3 & 0 & 0 \\
            0 & 0 & 12 & 6 \\
            0 & 0 & 6 & 4
        \end{pmatrix}
    \end{equation*}
    as $S\to\infty$. The corresponding scaled dilution of precision
    $\sqrt{S}\cdot\dop(\bm{\msf{G}})$ converges in probability to $\sqrt{22}\approx
    4.69$ as $S\to\infty$.
\end{theorem}

The proof of Theorem~\ref{thm:pr} is reported in Appendix~\ref{sec:proofs_thm1}.

Theorem~\ref{thm:pr} shows that, as the number of satellites grows, the dilution
of precision decreases as $\sqrt{22/S}$ with a stochastic variability (due to
the random satellite geometry) that is much smaller than $1/\sqrt{S}$. This
shows that increasing constellation size imparts two benefits. First, it
improves positioning performance with root-mean-squared (RMS) error decreasing
as the square root of the constellation size. Second, it reduces the variability
of the positioning performance because we most often have well distributed
satellites and consequently good satellite geometry.

The expression for the asymptotic covariance matrix $\bm{Q}$ in
Theorem~\ref{thm:pr} also shows that the vertical positioning RMS is
asymptotically twice as large as the horizontal positioning RMS (per dimension).
This is in line with empirical observations from smaller satellite
constellations~\cite{milbert09}. Further, asymptotically only the vertical
position and the clock bias estimation errors are correlated.

\begin{figure}
    \centering 
    \includegraphics{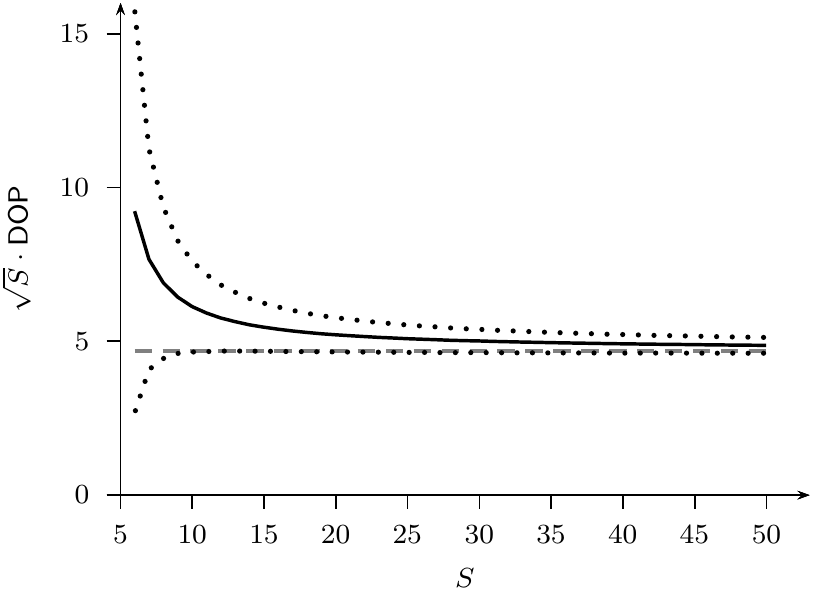}
    \caption{Dilution of precision as a function of number of satellites. The
        figure shows the expected value of $\sqrt{S}\cdot\dop(\bm{\msf{G}})$ (solid black
        line) plus/minus one standard deviation (dotted black lines) as a
        function of the number of satellites $S$. Also shown
        is the limiting value $\sqrt{22}$ from Theorem~\ref{thm:pr} (dashed gray line).}
    \label{fig:dop}
\end{figure}

Theorem~\ref{thm:pr} only provides asymptotic information about the behavior of
$\dop(\bm{\msf{G}})$ as $S$ increases. However, as Fig.~\ref{fig:dop} indicates,
the limiting behavior is already apparent for $S=20$ satellites.

\subsection{Carrier-Phase Positioning}
\label{sec:results_cp}

As mentioned earlier, dealing with the integer ambiguities present in the
carrier-phase measurements is one of the key challenges in successfully using
them for positioning. To see the potential value of the carrier-phase
measurements, assume for the moment that we knew the integer ambiguities
$\bm{m}$ exactly. A short computation shows that the ML estimate
of the parameter vector $\bm{w}$ is then given by
\begin{equation}
    \label{eq:cpmlgenie}
    (\bm{\msf{G}}^\T\bm{\msf{G}})^{-1}\bm{\msf{G}}^\T
    \Bigl(\frac{\sigma^{-2}}{\sigma^{-2}+\tilde{\sigma}^{-2}}\bm{\msf{y}}
    +\frac{\tilde{\sigma}^{-2}}{\sigma^{-2}+\tilde{\sigma}^{-2}}
    \bigl(\tilde{\bm{\msf{y}}}-\lambda\bm{m}\bigr)\Bigr).
\end{equation}
This estimate constructs a convex combination between the pseudo-range
measurements $\bm{\msf{y}}$ and the ambiguity-corrected carrier-phase
measurements $\tilde{\bm{\msf{y}}}-\lambda\bm{m}$. Since
$\tilde{\sigma}\ll\sigma$, the ambiguity-corrected carrier-phase measurements
have much higher weight in the convex combination than the pseudo-range
measurements. This estimator is again Gaussian with mean $\bm{w}$ (and therefore
unbiased) and with covariance matrix
\begin{equation*}
    \frac{1}{\sigma^{-2}+\tilde{\sigma}^{-2}}(\bm{\msf{G}}^\T\bm{\msf{G}})^{-1}.
\end{equation*}
Comparing this to~\eqref{eq:prcov}, we see that carrier-phase positioning with
known ambiguities is much more precise than pseudo-range only positioning. From
Theorem~\ref{thm:pr}, we also see that this covariance matrix scaled by $S$
converges in probability to  
\begin{equation}
    \label{eq:qgenie}
    \frac{1}{\sigma^{-2}+\tilde{\sigma}^{-2}}\bm{Q}
\end{equation}
as the number of satellites $S$ increases.

Of course, in reality we do not have direct access to the integer ambiguities
and instead need to estimate or resolve them from the measurements. The standard
approach for resolving the integer ambiguities consists of the following four
steps:\footnote{As long as the pseudo-range only estimate of the clock bias is
    not very precise, it is difficult to disentangle the effect of the integer
    ambiguities and the clock bias on the carrier-phase measurements. To
    alleviate this problem, the standard procedure is to resolve the
    \emph{differenced} ambiguities, i.e., $\msf{m}_s-\msf{m}_1$. However, for 
    the purposes of this paper, we can focus on the undifferenced ambiguities.}
\begin{enumerate}
    \item Find the \emph{float} estimate $\hat{\bm{\msf{m}}}$ of $\bm{m}$
        given the pseudo-range and carrier-phase measurements. This is the
        ML estimate ignoring the integer constraints.
    \item Compute the \emph{fixed} estimate $\breve{\bm{\msf{m}}}$ of $\bm{m}$
        by finding the closest (in the least-squares sense, taking into account
        the covariance of the float estimates) integer vector to
        $\hat{\bm{\msf{m}}}$.
    \item Validate the integer solution $\breve{\bm{\msf{m}}}$ by
        applying a statistical test that guarantees that the probability $\Pp(
        \breve{\bm{\msf{m}}} = \bm{m})$ is above some threshold close to one.
    \item Assuming validation was successful, estimate the receiver position and
        clock bias from the pseudo-range and carrier-phase measurements,
        treating the ambiguities as known and equal $\breve{\bm{\msf{m}}}$.
\end{enumerate}

As we will see next, this standard approach is unfortunately not appropriate for
the regime of large satellite constellations. In fact, as $S$ increases, correct
ambiguity resolution fails with probability approaching one.  To see this,
assume for the moment that a genie provides the correct value of the position
$\bm{x}$ and the clock bias $b$ to the receiver. Clearly, this knowledge can
only increase the probability of successful integer ambiguity resolution. The ML
(or, equivalently, the least-squares) integer estimate $\breve{\msf{m}}_s$ of
$m_s$ is then given by 
\begin{equation*}
    \breve{\msf{m}}_s \defeq \lfloor (\tilde{\msf{y}}_s + \bm{\msf{u}}_s^\T\bm{x}-b)/\lambda \rceil,
\end{equation*}
where $\lfloor\cdot\rceil$ denotes rounding to the closest integer. Now note
that
\begin{equation*}
    \breve{\msf{m}}_s = \lfloor m_s+\tilde{\sigma}\tilde{\msf{z}}_s/\lambda \rceil,
\end{equation*}
and therefore
\begin{equation*}
    \Pp(\breve{\msf{m}}_s = m_s)
    = 1-2\Phi\bigl(-\lambda/(2\tilde{\sigma})\bigr)
\end{equation*}
with $\Phi(\cdot)$ denoting the standard Gaussian cumulative distribution
function. Since the $\breve{\msf{m}}_s$ are independent, this implies that
\begin{equation}
    \label{eq:pcorr}
    \Pp(\breve{\bm{\msf{m}}} = \bm{m})
    = \Bigl(1-2\Phi\bigl(-\lambda/(2\tilde{\sigma})\bigr)\Bigr)^S
    \to 0
\end{equation}
as $S\to\infty$ provided that $\lambda/\tilde{\sigma}$ is finite. Thus, even
with the aid of the genie providing $\bm{x}$ and $b$, ambiguity resolution fails
with probability one as the number of satellites increases. Clearly, the same
conclusion holds without the aid of the genie.

From this discussion, we see that for a large number of satellites, we are
unable to correctly resolve all the ambiguities. A different approach is therefore
required. In order to avoid having to resolve the ambiguities, we will instead
treat them as a noise term and estimate the position and clock bias directly
from the pseudo-range and carrier-phase measurements. To this end, we place a
prior distribution on the integer ambiguities. Specifically, we assume in the
following that the ambiguities $\msf{m}_1, \dots, \msf{m}_S$ are \iid uniformly
distributed over the set $\{-M, -M+1, \dots, M-1, M\}$ with $M$ a fixed positive
integer. We will mainly be interested in scenarios where $M$ is large enough to
ensure that the resulting prior distribution on the ambiguities contains little
information. Specifically, this is the case when $M \gg \sigma/\lambda$.

\begin{figure}
    \centering 
    \includegraphics{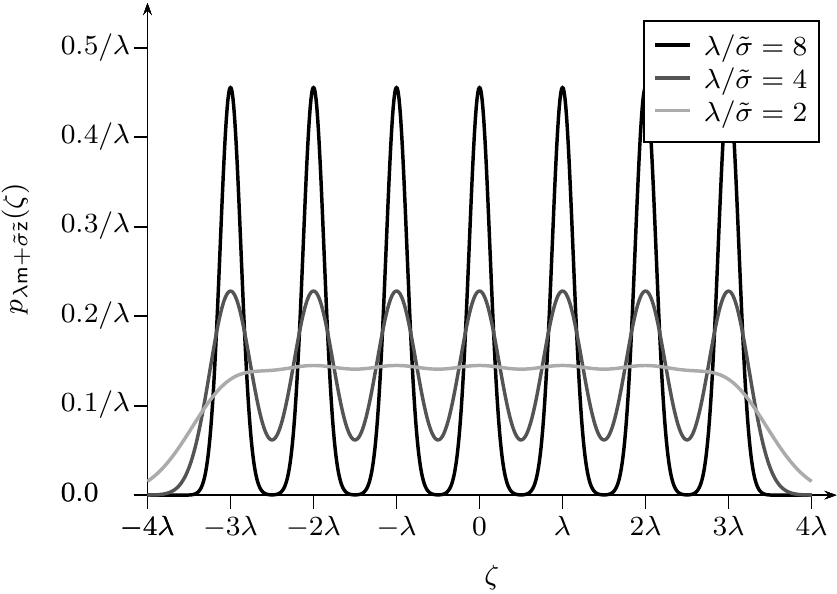}
    \caption{Probability density function
        $p_{\lambda\msf{m}+\tilde{\sigma}\tilde{\msf{z}}}$ of the combined carrier
        phase noise for different ratios of carrier wavelength to receiver noise
        standard deviation $\lambda/\tilde{\sigma}$. This combined noise includes
        the integer ambiguity, which is uniformly distributed on $\{-M, -M+1,
        \dots, M-1, M\}$. In the figure, $M=3$.}
    \label{fig:noisepdf}
\end{figure}

The probability density function of the combined carrier-phase noise (ambiguity
plus receiver noise) is shown in Fig.~\ref{fig:noisepdf}. The combined noise has
a mixture Gaussian density. For $\lambda/\tilde{\sigma}=8$, the peaks of the
mixture components are clearly distinguishable. On the other hand, for
$\lambda/\tilde{\sigma}=2$, they are virtually indistinguishable, and the
combined carrier phase noise distribution is almost uniform over the interval
$[-\lambda M, \lambda M]$.
 
Treating $\bm{\msf{m}}$ as noise, the ML estimator $\hat{\bm{\msf{w}}}$ of
$\bm{w}$ is shown in Appendix~\ref{sec:proofs_thm2} to be a solution of the
equation
\begin{align} 
    \label{eq:cpml}
    \hat{\bm{w}} 
    = (\bm{\msf{G}}^\T\bm{\msf{G}})^{-1}\bm{\msf{G}}^\T
    \Bigl(
    \frac{\sigma^{-2}}{\sigma^{-2}+\tilde{\sigma}^{-2}}\bm{\msf{y}}
    + \frac{\tilde{\sigma}^{-2}}{\sigma^{-2}+\tilde{\sigma}^{-2}}
    \bigl(
    \tilde{\bm{\msf{y}}} - \lambda \E_{\hat{\bm{w}}}(\bm{\msf{m}} \mid \tilde{\bm{\msf{y}}}, \bm{\msf{G}} )
    \bigl)
    \Bigr).
\end{align}
Here $\E_{\hat{\bm{w}}}(\cdot)$ denotes the expectation under the hypothesis
that the true parameter vector $\bm{w}$ takes the value $\hat{\bm{w}}$. The
conditional expectation in~\eqref{eq:cpml} has a simple closed-form expression
given by
\begin{equation}
    \label{eq:mmse}
    \E_{\hat{\bm{w}}}( \msf{m}_s \mid  \tilde{\bm{\msf{y}}},  \bm{\msf{G}})
    = \frac{\sum_{m=-M}^M m \exp\bigl(-\tfrac{1}{2\tilde{\sigma}^2}
    (\tilde{\msf{y}}_s - \bm{\msf{g}}_s^\T\hat{\bm{w}} - \lambda m)^2
    \bigr)}
    {\sum_{m=-M}^M \exp\bigl(-\tfrac{1}{2\tilde{\sigma}^2}
    (\tilde{\msf{y}}_s - \bm{\msf{g}}_s^\T\hat{\bm{w}} - \lambda m)^2
    \bigr)}
\end{equation}
and can be efficiently evaluated.

Note that the value of the expectation $\E_{\hat{\bm{w}}}(\bm{\msf{m}} \mid
\tilde{\bm{\msf{y}}}, \bm{\msf{G}} )$ depends itself on the value of
$\hat{\bm{w}}$. As a consequence, the ML estimator $\hat{\bm{\msf{w}}}$ is
implicitly defined as a solution of~\eqref{eq:cpml}.  In general, there will be
more than one solution $\hat{\bm{\msf{w}}}$ satisfying~\eqref{eq:cpml} (see
Fig.~\ref{fig:contour} below). To ensure uniqueness, we let $\hat{\bm{\msf{w}}}$
be the solution of \eqref{eq:cpml} that is closest (in Euclidean norm) to the
pseudo-range only estimator
$(\bm{\msf{G}}^\T\bm{\msf{G}})^{-1}\bm{\msf{G}}^\T\bm{\msf{y}}$.

Comparing~\eqref{eq:cpml} with~\eqref{eq:cpmlgenie}, we see that the vector
$\E_{\hat{\bm{w}}}\bigl(\bm{\msf{m}} \mid \tilde{\bm{\msf{y}}}, \bm{\msf{G}}
\bigr)$ can be interpreted as an estimate of the integer ambiguities
$\bm{\msf{m}}$. In fact, it is the minimum mean-squared error (MMSE) estimator
of the ambiguities. We emphasize that this MMSE estimator takes the integer
nature of the ambiguities into account, as can be seen from~\eqref{eq:mmse}.

The next theorem characterizes the asymptotic behavior of the estimator
$\hat{\bm{\msf{w}}}$ in \eqref{eq:cpml} as the number of satellites $S$
increases.

\begin{theorem} 
    \label{thm:cp}
    For every fixed $M > 0$, the estimator $\hat{\bm{\msf{w}}}$ is
    consistent, i.e., $\hat{\bm{\msf{w}}}$ converges to $\bm{w}$ in probability
    as $S\to\infty$.  Further, the scaled estimation error $\sqrt{S} \cdot
    (\hat{\bm{\msf{w}}} - \bm{w})$ converges in distribution to a Gaussian
    vector with mean zero and covariance matrix
    \begin{equation*}
        \frac{1}{\sigma^{-2} + h_M(\lambda/\tilde{\sigma})\cdot\tilde{\sigma}^{-2}} \bm{Q}
    \end{equation*}
    as $S\to\infty$, where the matrix $\bm{Q}$ was defined in Theorem
    \ref{thm:pr} and where the function $h_M(\cdot)$ is defined by~\eqref{eq:h}
    in Appendix~\ref{sec:proofs_thm2}.
\end{theorem}

\begin{figure}
    \centering 
    \includegraphics{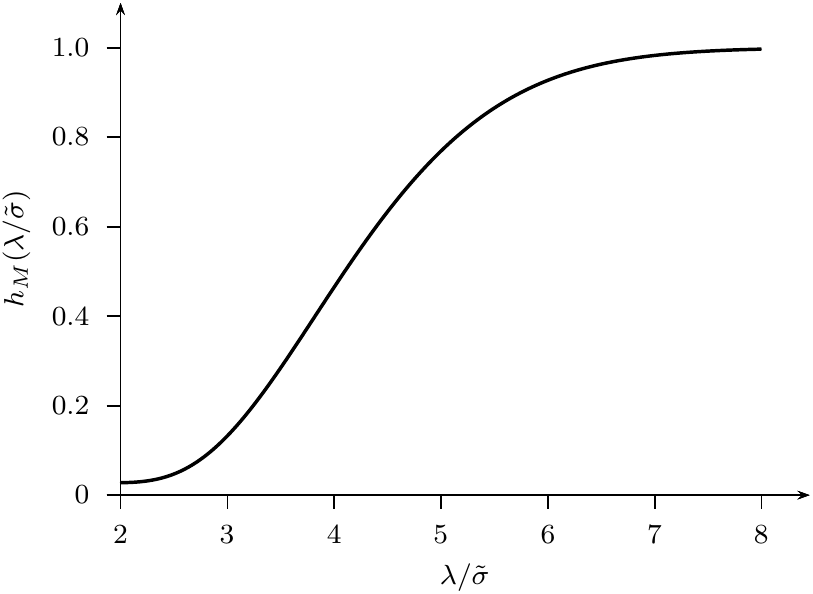}
    \caption{Factor $h_M(\lambda/\tilde{\sigma})$ from Theorem~\ref{thm:cp} as a
    function of $\lambda/\tilde{\sigma}$ for fixed value of $M=20$.}
    \label{fig:h}
\end{figure}

The proof of Theorem~\ref{thm:cp} is reported in Appendix~\ref{sec:proofs_thm2}.

Observe the similarity of the asymptotic covariance expression for carrier-phase
positioning in Theorem~\ref{thm:cp} with the ones for pseudo-range positioning
in Theorem~\ref{thm:pr} and for carrier-phase positioning with known ambiguities
in~\eqref{eq:qgenie}. Each of these expressions consist of the same matrix
$\bm{Q}$ pre-multiplied by a scalar factor. In order to illuminate the
connection between these scalar factors, we plot in Fig.~\ref{fig:h} the
function $h_M(\cdot)$ appearing in Theorem~\ref{thm:cp} for $M=20$. 
The figure shows that $h_M(\lambda/\tilde{\sigma})$ increases%
\footnote{For fixed and finite $M$, the function $h_M(a)$ is not
    increasing on the entire domain $a\in[0,\infty)$. In fact, $h_M(0)=1$ for
    all $M$, which can be seen by noting that for $\lambda = 0$ we always
    have $\lambda\msf{m}_s = 0$. However, for large enough $M$, the function
    $h_M(a)$ becomes increasing on the domain of interest. In particular,
    $h_M(a)$ is increasing on $a\in[2,\infty)$ for any fixed $M \geq 16$.} 
from $0$ to $1$ as a function of ratio $\lambda/\tilde{\sigma}$, which can be
interpreted as a (square-root) signal-to-noise ratio of the carrier-phase
signal.  

When this ratio is small, $h(\lambda/\tilde{\sigma})$ is close to $0$, and we
(approximately) recover the result of Theorem~\ref{thm:pr}. In this case, the
asymptotic covariance of the ML estimator is given by $\sigma^2\bm{Q}$, which
shows that the carrier-phase signal does not help.  On the other hand, when
$\lambda/\tilde{\sigma}$ increases, $h(\lambda/\tilde{\sigma})$ approaches the
value of $1$ (reaching this value approximately at $\lambda/\tilde{\sigma}=8$).
In this case, the asymptotic covariance of the ML estimator becomes
$\frac{1}{\sigma^{-2}+\tilde{\sigma}^{-2}}\bm{Q}$, which by~\eqref{eq:qgenie} is
the same as if we knew the integer ambiguities exactly.

Recall that typically $\tilde{\sigma}$ is significantly smaller than $\sigma$.
As a consequence, Theorem~\ref{thm:cp} implies that carrier-phase measurements
can yield substantial performance gains even if $h_M(\lambda/\tilde{\sigma})$ is
fairly small. In particular, we see from Fig.~\ref{fig:h} that carrier-phase
measurements are useful even if $\lambda/\tilde{\sigma}$ is below the value of
$4$. This contradicts the folklore rule of thumb that carrier-phase measurement
noise with a standard deviation larger than one quarter of the wavelength
renders those measurements unusable (see e.g.~\cite{petovello14, murrian16}).
Put differently, while it is difficult to resolve the integer ambiguities when
$\lambda/\tilde{\sigma}\leq 4$, the Bayesian approach adopted here (which does
not explicitly resolve the ambiguities) shows that the carrier-phase
measurements can still be beneficial in this regime. 

\begin{figure}
    \centering 
    \includegraphics{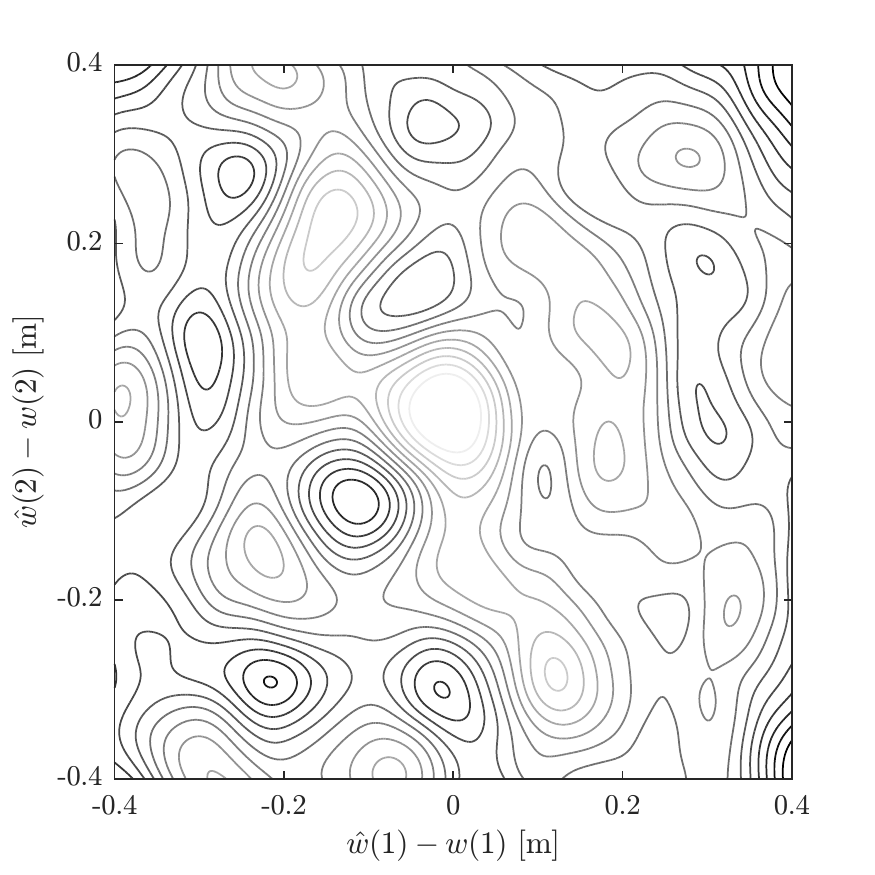}

    \caption{Contours (with brighter lines indicating larger values) of the
        log-likelihood as a function of $\hat{w}(1)$ and $\hat{w}(2)$. The
        remaining two components $\hat{w}(3)$ and $\hat{w}(4)$ are fixed to the
        true parameter values $w(3)$ and $w(4)$, respectively. In the figure,
        $\sigma=\SI{1}{\meter}$, $\lambda = \SI{0.19}{\meter}$,
        $\lambda/\tilde{\sigma} = 4$, $S=50$, and $M=20$.}

    \label{fig:contour}
\end{figure}

Our discussion so far has focused on the asymptotic behavior of carrier-phase
positioning. We next evaluate the performance of the Bayesian carrier-phase
positioning approach adopted in this paper for finite number $S$ of satellites.
Recall that the ML estimator $\hat{\bm{\msf{w}}}$ of $\bm{w}$ is a solution of
\eqref{eq:cpml}, which results from setting the derivative of the log-likelihood
to zero. As Fig.~\ref{fig:contour} shows, the log-likelihood function has
usually a fairly large number of local minima, maxima, and saddle points. Each
of those corresponds to a solution of \eqref{eq:cpml}. 

For our asymptotic analysis, we chose as estimator the solution of
\eqref{eq:cpml} closest to the pseudo-range only estimator. While this choice is
asymptotically optimal and works well for very large numbers of satellites ($S
\gg 100$), it unfortunately performs poorly for more realistic numbers of
satellites (say $S=50$). In this regime it is beneficial to instead choose the
value of $\hat{\bm{\msf{w}}}$ that directly maximizes the log-likelihood given
by~\eqref{eq:loglikelihood} in Appendix~\ref{sec:proofs_thm2}. This maximizer
can be found by running a global multi-start optimization procedure in a
neighbourhood of the pseudo-range only estimator.

\begin{figure}
    \centering 
    \includegraphics{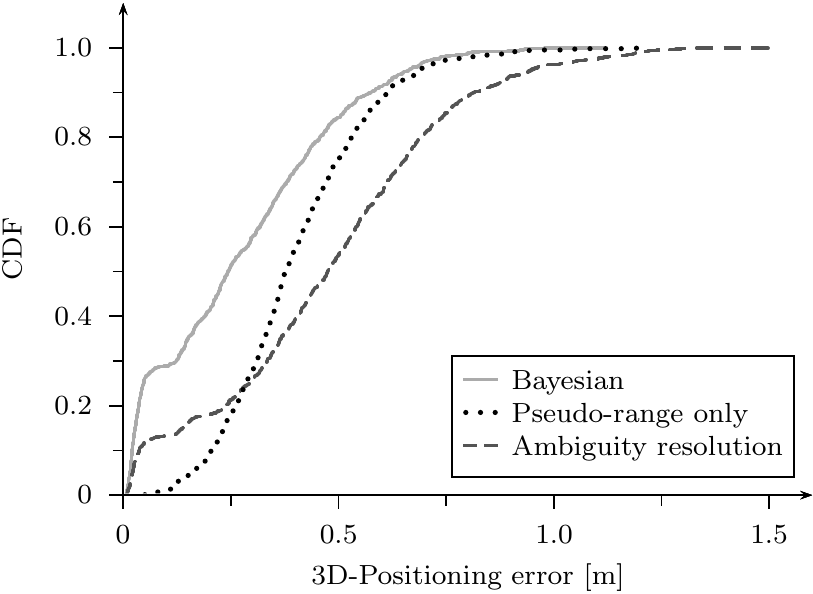}

    \caption{Error cumulative distribution function (CDF) for three different
        positioning approaches: Pseudo-range only positioning (dotted line),
        carrier-phase positioning using standard ambiguity resolution (dashed
        line), and carrier-phase positioning using the Bayesian approach
        adopted here (solid line). In the figure, $\sigma=\SI{1}{\meter}$, $\lambda =
        \SI{0.19}{\meter}$, $\lambda/\tilde{\sigma} = 4$, $S=50$, and $M=20$.}

    \label{fig:errorcdf}
\end{figure}

The performance of this approach is depicted in Fig.~\ref{fig:errorcdf}. The
figure also shows the performance of pseudo-range only positioning and of
carrier-phase positioning using the standard ambiguity resolution approach. The
ratio of carrier wavelength to carrier-phase noise standard deviation is set to
$\lambda/\tilde{\sigma} = 4$ and the number of satellites is $S=50$. Recall from
the discussion above that ambiguity resolution is considered difficult or
impossible in the regime $\lambda/\tilde{\sigma} = 4$. Indeed,
from~\eqref{eq:pcorr}, the probability of correctly resolving all the
ambiguities using the standard approach is less $\SI{10}{\percent}$. This small
probability of correct ambiguity resolution results in the poor performance of
the standard approach: As Fig.~\ref{fig:errorcdf} shows, carrier-phase
positioning using the standard ambiguity resolution approach performs worse than
pseudo-range only positioning about $\SI{70}{\percent}$ of the time. In
contrast, the Bayesian carrier-phase positioning approach adopted in this paper
results in a noticeable performance improvement compared to pseudo-range only
positioning.

\section{Conclusion}
\label{sec:conclusion}

Motivated by the ever increasing number of available satellites, we have studied
the problem of satellite positioning with large constellations. We have derived
the asymptotic behavior of both pseudo-range and carrier-phase positioning. For
carrier-phase positioning, we have argued that standard ambiguity resolution
fails for large number of satellites, and a Bayesian approach of treating those
ambiguities as additional receiver noise is more appropriate.

The results presented here raise several questions for follow-up work. First,
the maximization of the likelihood function for carrier-phase positioning using
a multi-start global search procedure is computationally quite demanding.
Devising algorithms to (approximately) solve this optimization problem
computationally more efficiently is therefore of interest. 

Second, while we have shown with an example that carrier-phase positioning
benefits from the Bayesian treatment of the ambiguities even for finite number
of satellites, the same example also indicates that the number of satellites
needs to be quite large for the asymptotic performance predictions to be
accurate. It would therefore be beneficial to have analytical performance
guarantees valid for smaller number of satellites.

\section*{Acknowledgment}

The authors thank Nicolas Macris and Victor Panaretos for their help regarding
the proof of Theorem~\ref{thm:cp}.

\appendix

\subsection{Proof of Theorem~\ref{thm:pr}}
\label{sec:proofs_thm1}

The matrix $\bm{\msf{G}}^\T\bm{\msf{G}}$ can be written as
\begin{equation*}
    \bm{\msf{G}}^\T\bm{\msf{G}}
    = \sum_{s=1}^{S} 
    \begin{pmatrix}
        -\bm{\msf{u}}_s \\
        1
    \end{pmatrix}
    \begin{pmatrix}
        -\bm{\msf{u}}_s^\T & 1
    \end{pmatrix}.
\end{equation*}
Observe that the matrices
\begin{equation*}
    \begin{pmatrix}
        -\bm{\msf{u}}_s \\
        1
    \end{pmatrix}
    \begin{pmatrix}
        -\bm{\msf{u}}_s^\T & 1
    \end{pmatrix}
\end{equation*}
are \iid as a function of $s$ (since the unit vectors $\bm{\msf{u}}_s$ are) and
have finite expected value. Hence, the weak law of large numbers applies and
shows that
\begin{equation}
    \label{eq:proof1_1}
    \frac{1}{S}\bm{\msf{G}}^\T\bm{\msf{G}}
    \overset{\Pp}{\underset{S \to \infty}{\to}} 
    \E\biggl(
    \begin{pmatrix}
        -\bm{\msf{u}}_1 \\
        1
    \end{pmatrix}
    \begin{pmatrix}
        -\bm{\msf{u}}_1^\T & 1
    \end{pmatrix}\biggr),
\end{equation}
where $\overset{\Pp}{\to}$ denotes convergence in probability, and where
$\E(\cdot)$ denotes expectation.

We next compute this expectation. We start with the diagonal terms. By
symmetry, we have 
\begin{equation*}
    \E(\msf{u}_{11}^2) = \E(\msf{u}_{12}^2) = \E(\msf{u}_{13}^2).
\end{equation*}
Further, since $\bm{\msf{u}}_1$ is a unit vector, we have
\begin{equation*}
    \E(\msf{u}_{11}^2) + \E(\msf{u}_{12}^2) + \E(\msf{u}_{13}^2) = 1.
\end{equation*}
Hence,
\begin{equation*}
    \E(\msf{u}_{11}^2) = \E(\msf{u}_{12}^2) = \E(\msf{u}_{13}^2) = 1/3.
\end{equation*}
The cross terms $\msf{u}_{1i}\msf{u}_{1j}$ for $i\neq j$ are easily seen to be
zero by symmetry, and similar for the cross terms $-1\cdot\msf{u}_{11}$ and
$-1\cdot\msf{u}_{12}$. It remains the cross term $-1\cdot\msf{u}_{13}$. A
straightforward calculation, making use of the standard expression for the area
of a spherical cap, shows that the marginal distribution of
$\msf{u}_{3i}$ is uniform on $[0,1]$. 
Hence,
\begin{equation*}
    \E(-1\cdot\msf{u}_{13}) = -1/2.
\end{equation*}
Together with~\eqref{eq:proof1_1}, this shows that
\begin{equation*}
    \frac{1}{S}\bm{\msf{G}}^\T\bm{\msf{G}}
    \overset{\Pp}{\underset{S \to \infty}{\to}} 
    \begin{pmatrix}
        1/3 & 0 & 0 & 0 \\
        0 & 1/3 & 0 & 0 \\
        0 & 0 & 1/3 & -1/2 \\
        0 & 0 & -1/2 & 1
    \end{pmatrix}.
\end{equation*}

Observe that this last matrix is invertible. Since the matrix inverse is
continuous, the continuous mapping theorem implies that
\begin{align*}
    \Bigl(\frac{1}{S}\bm{\msf{G}}^\T\bm{\msf{G}}\Bigr)^{-1}
    & \overset{\Pp}{\underset{S \to \infty}{\to}} 
    \begin{pmatrix}
        1/3 & 0 & 0 & 0 \\
        0 & 1/3 & 0 & 0 \\
        0 & 0 & 1/3 & -1/2 \\
        0 & 0 & -1/2 & 1
    \end{pmatrix}^{-1} \\
    & =
    \begin{pmatrix}
        3 & 0 & 0 & 0 \\
        0 & 3 & 0 & 0 \\
        0 & 0 & 12 & 6 \\
        0 & 0 & 6 & 4
    \end{pmatrix} \\
    & = \bm{Q}.
\end{align*}
A second application of the continuous mapping theorem further shows that
\begin{equation*}
    \sqrt{S}\cdot\dop(\bm{\msf{G}}) 
    \overset{\Pp}{\underset{S \to \infty}{\to}} \sqrt{\tr(\bm{Q})}
    = \sqrt{22},
\end{equation*}
as claimed. \hfill\IEEEQED

\subsection{Proof of Theorem~\ref{thm:cp}}
\label{sec:proofs_thm2}

Treating the integer ambiguities $\bm{\msf{m}}$ as noise uniformly distributed on
$\{-M,\ldots,M\}^S$ and considering the rows $\bm{\msf{g}}_s^\T$ of the matrix
$\bm{\msf{G}}$ as i.i.d.~observations, we obtain that the ML estimator
$\hat{\bm{\msf{w}}}$ should maximize the following likelihood function (given
the observables $\bm{\msf{y}}=\bm{y}$, $\tilde{\bm{\msf{y}}}=\tilde{\bm{y}}$ and
$\bm{\msf{G}}=\bm{G}$):
\begin{align*}
    p_{\bm{\msf{y}},\tilde{\bm{\msf{y}}},\bm{\msf{G}}} (\bm{y},\tilde{\bm{y}},\bm{G} \semcol \bm{w} )
    & = p_{\bm{\msf{G}}} (\bm{G}) \,  p_{\bm{\msf{y}},\tilde{\bm{\msf{y}}} \mid \bm{\msf{G}}} (\bm{y},\tilde{\bm{y}} \semcol \bm{w} \mid \bm{G} ) \\
    & = p_{\bm{\msf{G}}} (\bm{G}) \!\!\! \sum_{\bm{m} \in \{-M,\ldots,M\}^S} \!\!\! p_{\bm{\msf{m}}} (\bm{m}) \, p_{\bm{\msf{y}},\tilde{\bm{\msf{y}}} \mid \bm{\msf{G}},\bm{\msf{m}}} (\bm{y},\tilde{\bm{y}} \semcol \bm{w} \mid \bm{G}, \bm{m}) \\
    & = C \, \exp \Bigl( -\frac{1}{2\sigma^2} \, \norm{\bm{G} \bm{w} - \bm{y}}^2 \Bigr) 
    \sum_{\bm{m} \in \{-M,\ldots,M\}^S} \exp \Bigl( -\frac{1}{2\tilde{\sigma}^2} \, \norm{\bm{G} \bm{w} + \lambda \bm{m} - \tilde{\bm{y}}}^2 \Bigr)\\
    & = C \, \prod_{s=1}^S  \exp \Bigl( -\frac{1}{2\sigma^2} \, ( \bm{g}_s^\T \bm{w} - y_s)^2 \Bigr) 
    \sum_{m_s = -M}^M \exp \Bigl( -\frac{1}{2\tilde{\sigma}^2} \, (\bm{g}_s^\T \bm{w} + \lambda m_s - \tilde{y}_s )^2 \Bigr),
\end{align*}
where the normalization constant $C$ is given by
\begin{equation*}
    C \defeq \frac{p_{\bm{\msf{G}}} (\bm{G})}{(2 \pi \sigma \tilde{\sigma})^S \, (2M+1)^S}.
\end{equation*}
The corresponding log-likelihood function $L(\bm{y},\tilde{\bm{y}},\bm{G}
\semcol \bm{w} )$ is given by
\begin{equation}
    \label{eq:loglikelihood}
    L(\bm{y},\tilde{\bm{y}},\bm{G} \semcol \bm{w} ) 
    = \log( C ) + \sum_{s=1}^S \ell(y_s,\tilde{y}_s, \bm{g}_s \semcol \bm{w})
\end{equation}
with
\begin{align*}
    \ell & (y,\tilde{y}, \bm{g} \semcol \bm{w})
    \defeq -\frac{1}{2 \sigma^2} \, (\bm{g}^\T \bm{w} - y)^2
    + \log \Biggl( \sum_{m = -M}^M \exp \Bigl( -\frac{1}{2 \tilde{\sigma}^2} \, (\bm{g}^\T \bm{w} + \lambda m - \tilde{y})^2 \Bigr) \Biggr).
\end{align*}

By definition, the ML estimator $\hat{\bm{\msf{w}}}$ satisfies
\begin{equation}
    \label{eq:mldef}
    \frac{\partial L(\bm{\msf{y}},\tilde{\bm{\msf{y}}},\bm{\msf{G}} \semcol \bm{w})}{\partial w_i}
    \biggr|_{\bm{w} = \hat{\bm{\msf{w}}}}
    = \sum_{s=1}^S \frac{\partial \ell(\msf{y}_s,\tilde{\msf{y}}_s, \bm{\msf{g}}_s
    \semcol \bm{w})}{\partial w_i} \biggr|_{\bm{w} = \hat{\bm{\msf{w}}}} 
    = 0 
\end{equation}
for all $i\in\{1,\dots,4\}$.  The partial derivative of the summand in the
log-likelihood is
\begin{align}
    \label{eq:pd1}
    \frac{\partial \ell(y,\tilde{y}, \bm{g} \semcol \bm{w})}{\partial w_i}
    & = \frac{g_i}{\sigma^2} \, (y - \bm{g}^\T \bm{w}) + \frac{g_i}{\tilde{\sigma}^2} \, (\tilde{y} 
    - \bm{g}^\T \bm{w})
    - \frac{g_i \lambda}{\tilde{\sigma}^2} \; \frac{\sum_{m=-M}^M m \, f_{\tilde{y} - \bm{g}^\T \bm{w}}(m)}{\sum_{m=-M}^M f_{\tilde{y} - \bm{g}^\T \bm{w}}(m)} \nonumber \\
    & = \frac{g_i}{\sigma^2} \, (y - \bm{g}^\T \bm{w}) + \frac{g_i}{\tilde{\sigma}^2} \, (\tilde{y} - \bm{g}^\T \bm{w})
    - \frac{g_i \lambda}{\tilde{\sigma}^2} \, \bracket{m}_{\tilde{y} - \bm{g}^\T \bm{w}},
\end{align}
where
\begin{align}
    \label{eq:p_v}
    f_v(m) 
    & \defeq \exp \Bigl( -\frac{1}{2 \tilde{\sigma}^2} (\lambda m - v)^2 \Bigr) 
    \quad \text{for $v \in \R$} \notag\\
    \shortintertext{and}
    \bracket{m^k}_v 
    & \defeq \frac{\sum_{m=-M}^M m^k f_v(m)}{\sum_{m=-M}^M f_v(m)} 
    \quad \text{for $k\in\N$ and $v \in \R$}.\quad
\end{align}

Note that the above bracket notation is justified by the fact that 
\begin{equation*}
    \frac{f_v(m)}{\sum_{\tilde{m}=-M}^M f_v(\tilde{m})}
\end{equation*}
is a probability mass function on $m\in\{-M,\ldots,M\}$ for every $v \in \R$.
Given the particular form of $f_v(m)$, one may also interpret the above
bracket as the conditional expectation
\begin{equation*}
    \bracket{m^k}_v 
    = \E( \msf{m}_s^k \mid \lambda \msf{m}_s + \tilde{\sigma} \tilde{\msf{z}}_s = v ).
\end{equation*}
This, in turn, may be rewritten as
\begin{equation} 
    \label{eq:cond_exp}
    \bracket{m^k}_v 
    = \E( \msf{m}_s^k \mid \tilde{\msf{y}}_s - \bm{\msf{g}}_s^\T \bm{w} = v ).
\end{equation}
We will use several times in the following that
\begin{equation}
    \label{eq:dmv}
    \frac{\partial\bracket{m^k}_v}{\partial v}
    = \frac{\lambda}{\tilde{\sigma}^2}\bigl(
    \bracket{m^{k+1}}_v-\bracket{m^{k}}_v\bracket{m}_v \bigr),
\end{equation}
which can be verified after a short calculation.

Substituting \eqref{eq:pd1} and \eqref{eq:cond_exp} into~\eqref{eq:mldef}, we
obtain that the ML estimator satisfies the equation
\begin{align}
    \label{eq:mlcond}
    \sum_{s=1}^S \Bigl( \frac{\msf{g}_{si}}{\sigma^2} \, (\msf{y}_s - \bm{\msf{g}_s}^\T \hat{\bm{w}}) + \frac{\msf{g}_{si}}{\tilde{\sigma}^2} \, (\tilde{\msf{y}}_s - \bm{\msf{g}}_s^\T \hat{\bm{w}}) 
     - \frac{\msf{g}_{si} \lambda}{\tilde{\sigma}^2} \, \E_{\hat{\bm{w}}}( \msf{m}_s \mid \tilde{\msf{y}}_s - \bm{\msf{g}}_s^\T \hat{\bm{w}} ) \Bigr) = 0
\end{align}
for all $i\in\{1,\dots,4\}$, where $\E_{\hat{\bm{w}}}(\cdot)$ denotes
expectation under the hypothesis that the true parameter vector $\bm{w}$ equals
$\hat{\bm{w}}$. Observing that
\begin{equation*}
    \E_{\hat{\bm{w}}}( \msf{m}_s \mid \tilde{\msf{y}}_s - \bm{\msf{g}}_s^\T \hat{\bm{w}} ) = \E_{\hat{\bm{w}}}( \msf{m}_s \mid \tilde{\msf{y}}_s,  \bm{\msf{g}}_s) =\E_{\hat{\bm{w}}}( \msf{m}_s \mid \tilde{\bm{\msf{y}}},  \bm{\msf{G}})
\end{equation*}
by the independence of the observations under the hypothesis that $\bm{w} =
\hat{\bm{w}}$, \eqref{eq:mlcond} may be rewritten more compactly as
\begin{equation*}
    (\bm{\msf{G}}^\T\bm{\msf{G}}) \, (\sigma^{-2} + \tilde{\sigma}^{-2}) \, \hat{\bm{w}}
    = \bm{\msf{G}}^\T \left( \sigma^{-2} \, \bm{\msf{y}} + \tilde{\sigma}^{-2} \,\tilde{\bm{\msf{y}}} - \tilde{\sigma}^{-2} \, \lambda \, \E_{\hat{\bm{w}}}( \bm{\msf{m}} \mid  \tilde{\bm{\msf{y}}},  \bm{\msf{G}}) \right)
\end{equation*}
leading finally to \eqref{eq:cpml} in Section~\ref{sec:results_cp}. 

As pointed out earlier, this last equation may have multiple solutions, and we
choose $\hat{\bm{\msf{w}}}$ as the one closest to the pseudo-range only
estimator $(\bm{\msf{G}}^\T\bm{\msf{G}})^{-1}\bm{\msf{G}}^\T\bm{\msf{y}}$. As
the latter estimator is consistent (i.e., it converges in probability towards
the true parameter $\bm{w}$ as $S \to \infty$) by Theorem~\ref{thm:pr}, this
implies that $\hat{\bm{\msf{w}}}$ is also a consistent estimator
by~\cite[p.~453]{lehmann98}.

The asymptotic normality of the ML estimator follows
from~\cite[Theorem~5.4]{knight00}. For that theorem to apply, the following
seven conditions must be satisfied.

\emph{Condition 1: The parameter space is open.} Since the parameter space is
the whole $\R^4$, this is clearly the case.

\emph{Condition 2: The support of  $\exp\bigl(\ell(y,\tilde{y},\bm{g} \semcol
\bm{w})\bigr)$ does not depend on $\bm{w}$. } Clearly, the set
$\bigl\{(y,\tilde{y},\bm{g}) : \exp\bigl(\ell(y,\tilde{y},\bm{g} \semcol
\bm{w})\bigr) >0\bigr\}$ does not depend on $\bm{w}$, so that this condition
is satisfied.

\emph{Condition 3: The mapping $\bm{w} \mapsto \exp\bigl(\ell(y,\tilde{y},\bm{g} \semcol
\bm{w})\bigr)$ is three times continuously differentiable for
every $(y,\tilde{y},\bm{g})$.} This condition again clearly holds.

\emph{Condition 4: For every $i\in\{1,\dots, 4\}$, the equality
\begin{equation*}
    \E \Bigl( 
    \frac{\partial \ell(\msf{y}_s,\tilde{\msf{y}}_s, \bm{\msf{g}}_s \semcol \bm{w})}{\partial w_i} 
    \Bigr) =0
\end{equation*}
holds.} Observe first that
$\msf{y}_s - \bm{\msf{g}}_s^\T \bm{w} = \sigma \msf{z}_s$
and $\tilde{\msf{y}}_s - \bm{\msf{g}}_s^\T \bm{w} = \lambda \msf{m}_s +
\tilde{\sigma} \tilde{\msf{z}}_s$. From~\eqref{eq:pd1}, we then obtain
\begin{equation*}
    \frac{\partial \ell(\msf{y}_s,\tilde{\msf{y}}_s, \bm{\msf{g}}_s \semcol \bm{w})}{\partial w_i}
    = \frac{\msf{g}_{si}}{\sigma^2} \, (\sigma \msf{z}_s) 
    + \frac{\msf{g}_{si}}{\tilde{\sigma}^2} \, (\lambda \msf{m}_s + \tilde{\sigma} \tilde{\msf{z}}_s) 
    - \frac{\msf{g}_{si} \lambda}{\tilde{\sigma}^2} \bracket{m}_{\lambda \msf{m}_s +\tilde{\sigma} \tilde{\msf{z}}_s},
\end{equation*}
which is actually independent of $\bm{w}$. Using \eqref{eq:cond_exp}, we further
obtain
\begin{equation} 
    \label{eq:pd1bis}
    \frac{\partial \ell(\msf{y}_s,\tilde{\msf{y}}_s, \bm{\msf{g}}_s \semcol \bm{w})}{\partial w_i}
    = \msf{g}_{si} \, \Bigl( \frac{\msf{z}_s}{\sigma} + \frac{\tilde{\msf{z}}_s}{\tilde{\sigma}} + \frac{\lambda}{\tilde{\sigma}^2} \bigl( \msf{m}_s -
    \E( \msf{m}_s \mid \lambda \msf{m}_s + \tilde{\sigma} \tilde{\msf{z}}_s) \bigr) \Bigr).
\end{equation}
Therefore,
\begin{align*}
    \E \left( \frac{\partial \ell(\msf{y}_s,\tilde{\msf{y}}_s, \bm{\msf{g}}_s \semcol \bm{w})}{\partial w_i} \right)
    & = \E \biggl(\msf{g}_{si} \, \Bigl( \frac{\msf{z}_s}{\sigma} + \frac{\tilde{\msf{z}}_s}{\tilde{\sigma}} + \frac{\lambda}{\tilde{\sigma}^2} \, \bigl(\msf{m}_s - \E( \msf{m}_s \mid \lambda \msf{m}_s + \tilde{\sigma} \tilde{\msf{z}}_s)\bigr) \Bigr) \biggr)\\
    & =  \E(\msf{g}_{si}) \, \E \Bigl( \frac{\msf{z}_s}{\sigma} + \frac{\tilde{\msf{z}}_s}{\tilde{\sigma}} \Bigr) \\
    & = 0,
\end{align*}
where we have used the towering property of conditional expectation, that
$\msf{g}_{si}$, $\msf{m}_s$, $\msf{z}_s$, $\tilde{\msf{z}}_s$ are independent,
and that $\msf{z}_s, \tilde{\msf{z}}_s$ are centered. 

\emph{Condition 5: For every $\bm{w} \in \R^4$, the $4 \times 4$ matrix
\begin{equation*}
    \bm{I}(\bm{w})
    \defeq 
    \biggl( \E \Bigl( \frac{\partial \ell(\msf{y}_s,\tilde{\msf{y}}_s, \bm{\msf{g}}_s \semcol \bm{w})}{\partial w_i} \, \frac{\partial \ell(\msf{y}_s,\tilde{\msf{y}}_s, \bm{\msf{g}}_s \semcol \bm{w})}{\partial w_j} \Bigr) \biggr)_{i,j\in\{1,\ldots,4\}}
\end{equation*}
is positive-definite.} We start by deriving an explicit expression for the
matrix $\bm{I}(\bm{w})$. Reusing \eqref{eq:pd1bis}, we obtain
\begin{align}
    \label{eq:i1}
    I_{ij}(\bm{w})
    & = \E \left( \frac{\partial \ell(\msf{y}_s,\tilde{\msf{y}}_s, \bm{\msf{g}}_s \semcol \bm{w})}{\partial w_i} \, \frac{\partial \ell(\msf{y}_s,\tilde{\msf{y}}_s, \bm{\msf{g}}_s \semcol \bm{w})}{\partial w_j} \right) \nonumber\\
    & = \E \biggl( \msf{g}_{si} \, \Bigl( \frac{\msf{z}_s}{\sigma} + \frac{\tilde{\msf{z}}_s}{\tilde{\sigma}} + \frac{\lambda}{\tilde{\sigma}^2} \, \bigl(\msf{m}_s - \E( \msf{m}_s \mid \lambda \msf{m}_s + \tilde{\sigma} \tilde{\msf{z}}_s)\bigr) \Bigr) \nonumber\\
    & \hspace{1cm} \times \msf{g}_{sj} \, \Bigl( \frac{\msf{z}_s}{\sigma} + \frac{\tilde{\msf{z}}_s}{\tilde{\sigma}} + \frac{\lambda}{\tilde{\sigma}^2} \, \bigl(\msf{m}_s - \E( \msf{m}_s \mid \lambda \msf{m}_s + \tilde{\sigma} \tilde{\msf{z}}_s)\bigr) \Bigr) \biggr) \nonumber\\
    & = \E (\msf{g}_{si} \msf{g}_{sj})
    \biggl( \frac{1}{\sigma^2} 
    + \E \Bigl( \Bigl( \frac{\tilde{\msf{z}}_s}{\tilde{\sigma}} + \frac{\lambda}{\tilde{\sigma}^2} \, \bigl(\msf{m}_s - \E( \msf{m}_s \mid \lambda \msf{m}_s + \tilde{\sigma} \tilde{\msf{z}}_s)\bigr) \Bigr)^2 \Bigr) \biggr),
\end{align}
again by independence of $\bm{\msf{g}}_s$, $\msf{z}_s$, $\tilde{\msf{z}}_s$ and
$\msf{m}_s$, and using that $\msf{z}_s$ is centered. We can rewrite this last
equation in matrix form as
\begin{align}
    \label{eq:i2}
    \bm{I}(\bm{w})
    & = \bm{I} \\
    & = \E(\bm{\msf{g}}_s^\T\bm{\msf{g}}_s)
    \biggl( \frac{1}{\sigma^2} + \E \Bigl( \Bigl( \frac{\tilde{\msf{z}}_s}{\tilde{\sigma}} + \frac{\lambda}{\tilde{\sigma}^2} \, \bigl(\msf{m}_s - \E( \msf{m}_s \mid \lambda \msf{m}_s + \tilde{\sigma} \tilde{\msf{z}}_s)\bigr) \Bigr)^2 \Bigr) \biggr)  \\
    & = \bm{Q}^{-1}
    \biggl( \frac{1}{\sigma^2} + \E \Bigl( \Bigl( \frac{\tilde{\msf{z}}_s}{\tilde{\sigma}} + \frac{\lambda}{\tilde{\sigma}^2} \, \bigl(\msf{m}_s - \E( \msf{m}_s \mid \lambda \msf{m}_s + \tilde{\sigma} \tilde{\msf{z}}_s)\bigr) \Bigr)^2 \Bigr) \biggr),
\end{align}
where $\bm{Q}$ is the matrix given in Theorem~\ref{thm:pr}, and where the
notation $\bm{I}$ is used to indicate that the right-hand side does not depend
on $\bm{w}$. Since $\bm{Q}$ is positive definite, and since the scalar factor
multiplying this matrix is positive,  the matrix $\bm{I}(\bm{w})$ is
positive-definite for every $\bm{w} \in \R^4$ as required. 

For future reference, we further simplify the expression for $I_{ij}(\bm{w})$.
We can rewrite the second expectation in~\eqref{eq:i1} as
\begin{align}
    \label{eq:temp1}
    \E \Bigl( \Bigl( \frac{\tilde{\msf{z}}_s}{\tilde{\sigma}} & + \frac{\lambda}{\tilde{\sigma}^2} \, \bigl(\msf{m}_s - \E( \msf{m}_s \mid \lambda \msf{m}_s + \tilde{\sigma} \tilde{\msf{z}}_s)\bigr) \Bigr)^2 \Bigr) \nonumber \\
    & = \frac{1}{\tilde{\sigma}^2} + \frac{2 \lambda}{\tilde{\sigma}^3} \, \E \Bigl( \tilde{\msf{z}}_s \bigl( \msf{m}_s - \E( \msf{m}_s \mid \lambda \msf{m}_s + \tilde{\sigma} \tilde{\msf{z}}_s)\bigr)\Bigr)
    + \frac{\lambda^2}{\tilde{\sigma}^4} \, \E \Bigl(\bigl(\msf{m}_s - \E( \msf{m}_s \mid \lambda \msf{m}_s + \tilde{\sigma} \tilde{\msf{z}}_s)\bigr)^2 \Bigr) \nonumber \\
    & = \frac{1}{\tilde{\sigma}^2} - \frac{2 \lambda}{\tilde{\sigma}^3} \, \E \bigl( \tilde{\msf{z}}_s \, \E( \msf{m}_s \mid \lambda \msf{m}_s + \tilde{\sigma} \tilde{\msf{z}}_s)\bigr)
    + \frac{\lambda^2}{\tilde{\sigma}^4} \Bigl( \E  (\msf{m}_s^2) - \E \bigl( \E( \msf{m}_s \mid \lambda \msf{m}_s + \tilde{\sigma} \tilde{\msf{z}}_s)^2 \bigr) \Bigr),
\end{align}
using again the independence of $\tilde{\msf{z}}_s$ and $\msf{m}_s$, and using
the towering property of conditional expectation to conclude that
\begin{align*}
   \E\bigl( \msf{m}_s  \E( \msf{m}_s \mid \lambda \msf{m}_s + \tilde{\sigma} \tilde{\msf{z}}_s) \bigr)
   & = \E\Bigl(
   \E \bigl( \msf{m}_s  \E( \msf{m}_s \mid \lambda \msf{m}_s + \tilde{\sigma} \tilde{\msf{z}}_s) \bigm\vert \lambda \msf{m}_s + \tilde{\sigma} \tilde{\msf{z}}_s) \bigr) 
   \Bigr) \\
   & = \E \bigl( \E( \msf{m}_s \mid \lambda \msf{m}_s + \tilde{\sigma} \tilde{\msf{z}}_s)^2 \bigr).
\end{align*}
We next use the fact that for
$\tilde{\msf{z}}_s \sim \mc{N}(0,1)$ and for any continuously differentiable
function $F(\cdot)$ with polynomial growth, the integration by parts formula
gives
\begin{equation*}
    \E (\tilde{\msf{z}}_s \, F(\tilde{\msf{z}}_s)) = \E (F'(\tilde{\msf{z}}_s)).
\end{equation*}
Here, we would like to compute
\begin{equation*}
    \E \bigl(\tilde{\msf{z}}_s \, F( \lambda \msf{m}_s + \tilde{\sigma} \tilde{\msf{z}}_s)\bigr)
\end{equation*}
with
\begin{equation*}
    F(v) 
    \defeq \E( \msf{m}_s \mid \lambda \msf{m}_s + \tilde{\sigma} \tilde{\msf{z}}_s=v) 
    = \bracket{m}_v.
\end{equation*}
Using~\eqref{eq:dmv}, the derivative of $F(\cdot)$ is given by
\begin{equation*}
    F'(v) 
    = \frac{\partial}{\partial v} \bracket{m}_v 
    = \frac{\lambda}{\tilde{\sigma}^2} \left( \bracket{m^2}_v - \bracket{m}_v^2 \right).
\end{equation*}
Hence,
\begin{align*}
    \E\bigl( \tilde{\msf{z}}_s \E(\msf{m}_s \mid \lambda \msf{m}_s + \tilde{\sigma} \tilde{\msf{z}}_s) \bigr)
    & = \E \bigl(\tilde{\msf{z}}_s \, F( \lambda \msf{m}_s + \tilde{\sigma} \tilde{\msf{z}}_s)\bigr) \\
    & =  \tilde{\sigma} \, \E\bigl(F'(\lambda \msf{m}_s + \tilde{\sigma} \tilde{\msf{z}}_s)\bigr) \\
    & = \frac{\lambda}{\tilde{\sigma}} \, \E \left( \bracket{m^2}_{\lambda \msf{m}_s + \tilde{\sigma} \tilde{\msf{z}}_s} - \bracket{m}_{\lambda \msf{m}_s + \tilde{\sigma} \tilde{\msf{z}}_s}^2 \right)\\
    & =  \frac{\lambda}{\tilde{\sigma}} \, \E \Bigl(  \E\bigl( \msf{m}_s^2 \mid \lambda \msf{m}_s + \tilde{\sigma} \tilde{\msf{z}}_s\bigr) - \E( \msf{m}_s \mid \lambda \msf{m}_s + \tilde{\sigma} \tilde{\msf{z}}_s)^2 \Bigr)\\
    & = \frac{\lambda}{\tilde{\sigma}} \, \Bigl( \E (\msf{m}_s^2) -  \E \bigl(\E(\msf{m}_s \mid \lambda \msf{m}_s + \tilde{\sigma} \tilde{\msf{z}}_s)^2 \bigr) \Bigr).
\end{align*}
Substituting this into \eqref{eq:temp1} leads to
\begin{equation*}
    \E \biggl( \Bigr( \frac{\tilde{\msf{z}}_s}{\tilde{\sigma}} + \frac{\lambda}{\tilde{\sigma}^2} \, \bigl(\msf{m}_s - \E( \msf{m}_s \mid \lambda \msf{m}_s + \tilde{\sigma} \tilde{\msf{z}}_s)\bigr) \Bigr)^2 \biggr)
    =  \frac{1}{\tilde{\sigma}^2} -  \frac{\lambda^2}{\tilde{\sigma}^4} \Bigl(  \E \bigl( \msf{m}_s^2 \bigr) - \E \bigl( \E( \msf{m}_s \mid \lambda \msf{m}_s + \tilde{\sigma} \tilde{\msf{z}}_s)^2 \bigr) \Bigr)
\end{equation*}
and finally to
\begin{equation} 
    \label{eq:I}
    I_{ij}(\bm{w})
    = (\bm{Q}^{-1})_{ij} \biggl( \frac{1}{\sigma^2} + \frac{1}{\tilde{\sigma}^2}
    - \frac{\lambda^2}{\tilde{\sigma}^4} \Bigl( \E(\msf{m}_s^2) - \E \bigl( \E( \msf{m}_s \mid \lambda \msf{m}_s + \tilde{\sigma} \tilde{\msf{z}}_s)^2 \bigr) \Bigr) \biggr). \quad
\end{equation}

\emph{Condition 6: For every $\bm{w} \in \R^4$, the $4 \times 4$
matrix $\bm{J}(\bm{w})$ defined as
\begin{equation*}
    \bm{J}(\bm{w}) \defeq 
    \biggl( \E \Bigl( - \frac{\partial^2 \ell(\msf{y}_s,\tilde{\msf{y}}_s, \bm{\msf{g}}_s \semcol \bm{w})}{\partial w_i \partial w_j} \Bigr) \biggr)_{i,j\in\{1,\ldots,4\}}
\end{equation*}
is positive-definite.} It turns out in our case that $\bm{J}(\bm{w}) =
\bm{I}(\bm{w})$ for every $\bm{w} \in \R^4$. Since $\bm{I}(\bm{w})$ was already
shown to be positive definite when verifying Condition~5, this implies that
Condition~6 holds. To prove this equality, we compute the second-order partial
derivatives. Starting from~\eqref{eq:pd1} and using~\eqref{eq:dmv}, we obtain
\begin{equation}
    \label{eq:pd2}
    \frac{\partial^2 \ell(y,\tilde{y}, \bm{g} \semcol \bm{w})}{\partial w_i \partial w_j}
    = g_i \, g_j \Bigl( -\frac{1}{\sigma^2} - \frac{1}{\tilde{\sigma}^2} + \frac{\lambda^2}{\tilde{\sigma}^4} \, \bigl( \bracket{m^2}_{\tilde{y} - \bm{g}^\T \bm{w}} - \bracket{m}_{\tilde{y} - \bm{g}^\T \bm{w}}^2 \bigr) \Bigr).
\end{equation}
Using again that $\tilde{\msf{y}}_s - \bm{\msf{g}}_s^\T \bm{w} = \lambda \msf{m}_s +
\tilde{\sigma} \tilde{\msf{z}}_s$, we have
\begin{equation*}
    - \frac{\partial^2 \ell(\msf{y}_s,\tilde{\msf{y}}_s, \bm{\msf{g}}_s \semcol \bm{w})}{\partial w_i \partial w_j}
    = \msf{g}_{si} \, \msf{g}_{sj} \Bigl( \frac{1}{\sigma^2} + \frac{1}{\tilde{\sigma}^2} - \frac{\lambda^2}{\tilde{\sigma}^4} \, \left( \bracket{m^2}_{\lambda \msf{m}_s + \tilde{\sigma} \tilde{z}_s} - \bracket{m}_{\lambda \msf{m}_s + \tilde{\sigma} \tilde{z}_s}^2 \right) \Bigr).
\end{equation*}
Taking expectation and comparing to \eqref{eq:I} leads to
\begin{equation*}
    J_{ij}(\bm{w})
    = (\bm{Q}^{-1})_{ij} \, \biggl( \frac{1}{\sigma^2} + \frac{1}{\tilde{\sigma}^2} 
    -  \frac{\lambda^2}{\tilde{\sigma}^4} \Bigl( \E(\msf{m}_s^2) - \E \bigl( \E( \msf{m}_s \mid \lambda \msf{m}_s + \tilde{\sigma} \tilde{\msf{z}}_s)^2 \bigr) \Bigr) \biggr) 
    = I_{ij}(\bm{w}),
\end{equation*}
proving the claim.

\emph{Condition 7: For every $\bm{w}\in\R^4$, $\delta>0$, and 
$i, j, k \in\{1,\dots, 4\}$, there exists $D_{ijk}(\msf{y}_s,
\tilde{\msf{y}}_s, \bm{\msf{g}}_s)$ satisfying
\begin{equation*}
    \E \abs{D_{ijk}(\msf{y}_s,\tilde{\msf{y}}_s, \bm{\msf{g}}_s)} < + \infty
\end{equation*}
such that
\begin{equation*}
    \Bigl| \frac{\partial^3 \ell(\msf{y}_s,\tilde{\msf{y}}_s, \bm{\msf{g}}_s \semcol \bm{w})}{\partial w_i \partial w_j \partial w_k} \Bigr|\;\biggr|_{\bm{w}=\bar{\bm{w}}}
    \le D_{ijk}(\msf{y}_s,\tilde{\msf{y}}_s, \bm{\msf{g}}_s)
\end{equation*}
holds for every $\norm{\bar{\bm{w}} - \bm{w}} \le \delta$.} Starting
from~\eqref{eq:pd2} and using~\eqref{eq:dmv}, the third partial derivatives can
be calculated as
\begin{equation} 
    \label{eq:pd3}
    \frac{\partial^3 \ell(y,\tilde{y}, \bm{g} \semcol \bm{w})}{\partial w_i \partial w_j \partial w_k}
    = - g_i \, g_j \, g_k \; \frac{\lambda^3}{\tilde{\sigma}^6} \, \Bigl( \bracket{m^3}_{\tilde{y} - \bm{g}^\T \bm{w}} - 3 \bracket{m^2}_{\tilde{y} - \bm{g}^\T \bm{w}} \, \bracket{m}_{\tilde{y} - \bm{g}^\T \bm{w}}
    + 2 \bracket{m}_{\tilde{y} - \bm{g}^\T \bm{w}}^3 \Bigr).
\end{equation}
We therefore have that
\begin{align*}
    \Bigl| \frac{\partial^3 \ell(\msf{y}_s,\tilde{\msf{y}}_s, \bm{\msf{g}}_s \semcol \bm{w})}{\partial w_i \partial w_j \partial w_k} \Bigr|\;\biggr|_{\bm{w}=\bar{\bm{w}}}
    & \le \abs{\msf{g}_{si} \, \msf{g}_{sj} \, \msf{g}_{sk}} \; \frac{\lambda^3}{\tilde{\sigma}^6} \; 6 M^3 \\
    & \le \frac{\lambda^3}{\tilde{\sigma}^6} \; 6 M^3,
\end{align*}
which is independent of $\bar{\bm{w}}$ and clearly integrable.

Under Conditions~1--7, \cite[Theorem~5.4]{knight00} states that
$\sqrt{S}\,(\hat{\bm{\msf{w}}} - \bm{w})$ converges in distribution to a
centered Gaussian random vector with covariance matrix
$\bm{J}(\bm{w})^{-1}\bm{I}(\bm{w})\bm{J}(\bm{w})^{-1} = \bm{J}(\bm{w})^{-1} =
\bm{I}^{-1}$ as $S\to\infty$.\footnote{Strictly speaking,
    \cite[Theorem~5.4]{knight00} is only stated for \emph{scalar}-valued \iid
    observations. The conclusion of the theorem remains however valid for
    \emph{vector}-valued \iid observations, provided that the likelihood
    function derived from the vector-valued observations $(\msf{y}_s,
    \tilde{\msf{y}}_s, \bm{\msf{g}}_s)$ verifies Conditions~1--7.}
 
What remains to be computed is a more explicit expression for the scalar factor
\begin{equation*}
    \frac{1}{\sigma^2} + \frac{1}{\tilde{\sigma}^2} -  \frac{\lambda^2}{\tilde{\sigma}^4} \left(  \E \left( \msf{m}_s^2 \right) - \E \left( \E( \msf{m}_s \mid \lambda \msf{m}_s + \tilde{\sigma} \tilde{\msf{z}}_s)^2 \right) \right).
\end{equation*}
multiplying the matrix $\bm{Q}^{-1}$ in the expression \eqref{eq:I} for
$\bm{I}$. To this end, recall from~\eqref{eq:cond_exp} that
\begin{equation*}
    \E \left( \msf{m}_s^2 \right) - \E \left( \E( \msf{m}_s \mid \lambda \msf{m}_s + \tilde{\sigma} \tilde{\msf{z}}_s)^2 \right)
    = \E \bigl( \bracket{m^2}_{\lambda \msf{m}_s + \tilde{\sigma} \tilde{\msf{z}}_s} - \bracket{m}_{\lambda \msf{m}_s + \tilde{\sigma} \tilde{\msf{z}}_s}^2 \bigr)
\end{equation*}
and from~\eqref{eq:p_v} that
\begin{align*}
    \bracket{m^k}_v 
    & = \frac{\sum_{m=-M}^M m^k \, f_v(m)}{\sum_{m=-M}^M f_v(m)} \\
    & =  \frac{\sum_{m=-M}^M m^k \, \exp \left( - \frac{1}{2\tilde{\sigma}^2} (\lambda m - v)^2 \right)}{\sum_{m=-M}^M \exp \left( - \frac{1}{2\tilde{\sigma}^2} (\lambda m - v)^2 \right)}.
\end{align*}
Thus,
\begin{align*}
    \bracket{m^k}_{\lambda \msf{m}_s + \tilde{\sigma} \tilde{\msf{z}}_s} 
    & =  \frac{\sum_{m=-M}^M m^k \, \exp\Bigl(-\frac{1}{2}\bigl(\frac{\lambda}{\tilde{\sigma}}(m-\msf{m}_s)-\tilde{\msf{z}}_s \bigr)^2 \Bigr)}
    {\sum_{m=-M}^M \exp\Bigl(-\frac{1}{2}\bigl(\frac{\lambda}{\tilde{\sigma}}(m-\msf{m}_s)-\tilde{\msf{z}}_s \bigr)^2 \Bigr)} \\
    & \defeq H_k(\lambda/\tilde{\sigma}, \msf{m}_s, \tilde{\msf{z}}_s),
\end{align*}
which depends on the parameters $\lambda$ and $\tilde{\sigma}$ only through
their ratio $\lambda/\tilde{\sigma}$. Finally, we obtain
\begin{equation*}
    \frac{1}{\sigma^2} + \frac{1}{\tilde{\sigma}^2} - \frac{\lambda^2}{\tilde{\sigma}^4} \left(  \E \left( \msf{m}_s^2 \right) - \E \left( \E( \msf{m}_s \mid \lambda \msf{m}_s + \tilde{\sigma} \tilde{\msf{z}}_s)^2 \right) \right)
    = \sigma^{-2} + \tilde{\sigma}^{-2}h_M(\lambda/\tilde{\sigma})
\end{equation*}
with
\begin{equation}
    \label{eq:h}
    h_M(\lambda/\tilde{\sigma})
    \defeq 1 -  \frac{\lambda^2}{\tilde{\sigma}^2} \, \E \left( H_2( \lambda/\tilde{\sigma}, \msf{m}_s, \tilde{\msf{z}}_s) - H_1(\lambda/\tilde{\sigma}, \msf{m}_s, \tilde{\msf{z}}_s)^2 \right),
\end{equation}
where we have explicitly indicated the dependence of $h_M(\cdot)$ on the support
of the prior governed by $M$. Substituting this expression into~\eqref{eq:I}
shows that the inverse of the scaled asymptotic covariance matrix is
\begin{equation*}
    \bm{I} = \bigl(\sigma^{-2} + \tilde{\sigma}^{-2}h_M(\lambda/\tilde{\sigma})\bigr)\bm{Q}^{-1},
\end{equation*}
which can be rearranged as in the statement of the theorem.  This concludes the
proof. \hfill \IEEEQED

\end{document}